# Ways of Listening and Modes of Being: Electroacoustic Auditory Display

Paul Vickers

Northumbria University,

Dept. of Computer Science & Digital Technologies,
Pandon Building, Camden St.,
Newcastle upon Tyne NE2 1XE, UK

paul.vickers@northumbria.ac.uk

## Abstract

Auditory display is concerned with the use of non-speech sound to communicate information. If the term seems at first oxymoronic, then consider auditory display as an activity of *perceptualization*, that is, the process of making perceptible to humans aspects or features of a given data set or system. Most commonly this is done using visual representations (which process we call visualization) but it is not limited to the visual channel and recent years have witnessed the increased use of auditory representations in the production of tools for exploring data. By way of semiotics and an aesthetic perspective shift this article posits that auditory display may be considered a form of organized sound and explores the listening experience in this context.

## 1 Introduction

> Technology is a way of revealing.
> Martin Heidegger (Heidegger 1955)

Since Sara Bly's landmark thesis (Bly 1982) which showed how non-speech sound could be successfully used to communicate specific information a field of research known as auditory display has developed which concerns itself with using sound to represent data or system status. Far from being an oxymoron auditory display is simply a representational process in which sound is the representational medium the purpose of which is to 'help a user monitor and comprehend whatever it is that the sound output represents' (Kramer 1994b: 1). It is, perhaps, best grasped when compared to its visual counterpart, visualization, in which data or information is mapped to visual tropes to allow inspection and meaning making to take place.[1] So, where visualization would help a marketing manager to understand sales data by presenting them in chart form, auditory display might map the data to a series of tones in which a higher frequency denotes higher sales.

---

[1] I am thinking more of information visualization here, but scientific visualization also applies.

The term visualization generally refers to the processes that use visual display techniques to render data sets perceptible. However, visualization can also be used to describe the rendering and understanding of data via any modality including sound (auditory image). To avoid ambiguity, the term visualization is taken in this article to mean only those representational techniques that employ the visual channel. Perceptualization, then, is the higher level category that embraces representational practices across all the senses, i.e., sight, sound, touch, smell, and taste.

Auditory display practice embraces a wide variety of sonic presentation techniques from highly abstract tonal musical structures, through concrete recordings and soundscapes, to tones generated directly from the data themselves. Because auditory display is a representational process it is inherently a semiotic activity. This article explores the world of auditory display in terms of the listening experience, drawing parallels with musical and sonic art practice and putting it into a semiotic framework. Section 2 gives some background to the field of auditory display and its main categories. In Section 3 the semotic dimension is explored in relation to C.S. Peirce's three modes (or categories) of being. Section 4 shows how auditory display practice may be discussed in terms of organized sound and electroacoustic music. Section 5 then relates auditory display to different ways of listening, with an emphasis on Schaeffer's *quatre écoutes*. Finally some reflections and conclusions are offered.

## 2 What is auditory display?

Auditory display, then, is a family of representational processes in which loud speakers (or headphones) are used as information displays instead of visual display units. Whilst earlier examples can be found, the discipline tends to identify Sara Bly's 1982 Ph.D. thesis as a keystone. Ten years later saw the formation of the International Community for Auditory Display (ICAD) and its associated conference series.[2] The proceedings of that first International Conference on Auditory Display were expanded and reworked to become the seminal reference volume *Auditory Display* (Kramer 1994a). For a recent and up-to-date general introduction to the field of auditory display in all its variety see Hermann, Hunt, and Neuhoff (Hermann, Hunt, & Neuhoff 2011).[3] Like visualization, auditory display is at its heart a representational technique and an activity of perceptualization. That is, its goal is to make perceptible some data set or other attributes of a system or entity, knowledge of which will enable meaning making to take place.[4] Auditory display technology is thus 'a way of revealing' (Heidegger 1955).

---

[2] http://www.icad.org
[3] The Sonification Handbook is available to read and download free of charge at http://sonification.de/handbook/.
[4] Any discussion of making something perceptible via the senses necessarily touches upon the realm of aesthetics, a field which, at its core, concerns itself with sensuous perception. For a discussion of the relationship of auditory display to aesthetic practice see Barrass and Vickers (Barrass & Vickers 2011).

## 2.1 Sonification and Audification

Whilst we talk about auditory displays in the large the bulk of what falls under the auditory display umbrella can be classed as either sonification or audification. Sonification may be defined as:

> …the use of non-speech audio to convey information. More specifically, sonification is the transformation of data relations into perceived relations in an acoustic signal for the purposes of facilitating communication or interpretation (Kramer et al. 1999).

Where sonification uses data as the control input to some sound generating mechanism, audification is a much more direct representational scheme in which data samples are played back directly as sound (sometimes after some scaling to put them into the audible frequency range). For example, Chris Hayward's audio seismograms (Hayward 1994) are audifications because the data from seismic measuring equipment were scaled into an audible frequency range and played back through loud speakers.[5]

AudioObject1.mp3: Example of audio seismograms
http://sonification.de/handbook/media/chapter12/SHB-S12.16.mp3

See also Dombois' discussion of audification for seismology and explosion data (Dombois 2002).

# 3 Semiosis and Intended Meaning

The goal of auditory display is to make perceptible to the senses some aspect of a data set or a system so that the mind can examine that sensory input, draw inferences about it, and so gain insight into the subject of the display. The auditory display thus functions as a sign in the semiotic sense.

## 3.1 Semiotics and Modes of Being

Semiotics is the study of signs and sign systems, their creation and interpretation. Signs can be images, words, sounds, smells, etc. that have no inherent meaning but which become signs when we attribute meaning to them (Chandler 2007). Thus, a sign stands for something beyond itself. Modern semiotics is drawn from the work of two principal figures, the Swiss linguist Ferdinand de Saussure and the American scientist, logician, and philosopher Charles Sanders Peirce. In Saussure's structuralist view the sign system is a dyadic relationship between a *signifier* (typically a sound pattern) and a *signified* (the concept evoked by the signifier) both of which are non-material mental constructs rather than material objects (Chandler 2007) though modern semiotics does admit material form for the signifier (e.g., road signs, printed words, etc.). For example, the symbol /cat/ is a signifier for the concept of the thing we know as a cat. The resultant sign is a link between the sound pattern and the concept. It should be noted that Saussure explicitly excluded referent objects for the sign; Saussure's signs were arbitrary notations for referring to mental concepts already present in the mind of the addressee.

---

[5] The process was actually slightly more involved than that because seismic data has periodicities measured in days (and longer) and so to play back as audio requires some work. The data were first pre-processed with 'amplitude scaling, DC removal, and interpolation and then frequency-doubled, time compressed, and amplitude scaled (using automatic gain control) until they lay in the human audible range' (Vickers & Hogg 2006: 4).

Peirce considered such a dyadic relationship to be insufficient and he proposed a semiotic scheme in which a referent object is represented by a *representamen* which evokes in the mind of the consumer of the sign its meaning which he calls the *interpretant*. We can see that the interpretant loosely corresponds to Saussure's signified and the representamen to the signifier. This may be shown diagrammatically as

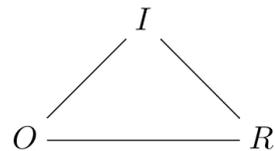

where *O*, *R*, and *I* correspond to the object, representamen, and interpretant respectively. Peirce was concerned with seeking to explain the meaning of signs with logical laws and proposed three *predicaments*, that is, modes of being (or 'categories' to use Aristotle's terminology) that could be used in the explanation of sign relationships:

1. Firstness deals with pure being. It is the existence of an object without reference to anything else; it is the unexperienced object.

2. Secondness describes pure relationship. Secondness is an encounter with firstness that excludes any thoughts or conceptions about the object.

3. Thirdness is the mediation of the relationship (secondness). Thirdness is the ideas and mental perceptions we have that relate firstness with secondness.

Thirdness, then, is the mode of being of signs in that signs mediate relationships between their objects and their interpretants. As well as introducing a referent object Peirce's semiotics also admits a ground for the relationship. The ground is an idea or principle which determines how the sign represents its object. It is thus a context in the Derridean sense. Peirce's semiotics, then, sits more comfortably in a post-structuralist world; the meaning of the sign (the interpretant) is constructed (in the light of the ground) from the representamen in the mind of the sign's addressee.[6]

Peirce's semiotics also addressed a criticism of the Saussurean school that Saussure's semiology considered there to be an absolute resolution to the process of interpreting a sign. In semiotics meaning arises through a process of differentiation: /this/ means this and /that/ means that because /this/ and /that/ are different signs and Saussure asserted that the process of differentiation is ultimately halted when the *transcendental signified* is reached.[7] Peirce, on the other hand, saw the interpretant evoked in the mind by a representamen as itself the representamen for a further round of signification. For Peirce this became a never-ending cycle of representation and interpretation. Derrida coined his word *différrance* to encapsulate the twin processes of differentiation and deferral which together make up this infinite regress of meaning making. For Derrida the meaning of any word in a text is modified by words that come later in the text, earlier in the text, and even outside the text, leading to the conclusion that there is nothing outside the text, that is, context is king and structuralism does not tell the whole story. However, whilst Peirce's signs may well admit an infinite cycle of signification and interpretation, the goal of auditory display is to arrive at some level of understanding of the data or phenomenon being represented. So, there needs to be, if not a transcendental interpretant, at least a point in the cycle at which we can say 'I see', or 'I understand'.

---

[6] Vickers, Faith, and Rossiter (Vickers, Faith, & Rossiter 2012) have shown how such a post-structuralist view can be combined with the algebraic tools of category theory to describe and understand with mathematical precision the process of visualization (and hence, by extension, auditory display).

Auditory display, then, is a semiotic activity in which a given data set (object) is represented by a sound signal (representamen) resulting in an awareness and understanding (interpretant) of that object in the mind of the listener. Furthermore, the relationship between the listener and the object is mediated by the auditory display and the ground upon which it sits. The existence of the ground is a necessary precondition for the auditory display to be successful as it constitutes a shared understanding between the auditory display's designer and its listener of the rules of the sonic representation (e.g., knowing that pitch represents stock price, timbre represents stock type, etc.). Having established that auditory display can be viewed as a Peircean (post-structuralist) semiotic relationship, we can now see how to understand auditory display in terms of organized sound.

## 4 Auditory Display as Organized Sound

Auditory displays are sonic representations or portrayals of objective data. How then may auditory display be considered in terms of music and organized sound? Browsing through the proceedings of the International Conference on Auditory Display one finds that auditory display researchers have employed a large number of semiotic coding schemes the significations of which have been appropriated for data perceptualization. From the language of tonal music one finds major- and minor-mode motifs being used to denote the Boolean values of *true* and *false* respectively; a return to a tonal centre denotes resolution, and so forth (e.g., the CAITLIN programming language sonification system (Vickers & Alty 2005)). A lot of sonifications are primitive data-to-equal-tempered-pitch mappings rendered (via the MIDI protocol) on a cheap synthesizer playing back tones from a General MIDI sound set. At the other end of scale can be found auditory displays that use the techniques of soundscape design (e.g., Gilfix and Crouch's PEEP system for sonifying network traffic (Gilfix & Couch 2000)). One can also find the influence of *musique concrète* and electroacoustic music, non-tonal and atonal schemes, and even the Futurists' influence can be found in the design of a web server sonification system (see Barra et al. 2002).

---

[7] Of course, /this/ and /that/ can sometimes both denote the same thing. Frege gave the example of the morning star and evening star. The terms have different intensional (connotative) meanings (a star that is seen in the morning and one that is seen in the evening) but they possess the same extension, or denotation, that is, the planet Venus (Sowa 2000: 99).

Emmerson (Emmerson 1986) devised the *language grid* as a device for categorizing electroacoustic music into a nine-sector two-dimensional space. The vertical axis indicates the music's mode of syntactical abstraction whilst the horizontal axis distinguishes between the use of mimetic reference and aural discourse.[8] Figure 1 (Vickers 2006) shows an elaborated version of the language grid in which sectors 7, 8, and 9 have been shaded to show the area of the grid predominantly occupied by auditory display. Popular forms of music, those based around pitch relations and employing harmony (tonal and atonal), because they are, by definition, abstract are located predominantly in sectors 1, 4, and 7 of the language grid (with absolute music on the left and programme music occupying the centre and right).

|                                         | 1 | 4 | 7 |
|---|---|---|---|
| Abstract syntax                         | 1 | 4 | 7 |
| Combination of abstract and abstracted syntax | 2 | 5 | 8 |
| Abstracted syntax                       | 3 | 6 | 9 |
|                                         | I: Aural discourse dominant | II: Combination of aural and mimetic discourse | III: Mimetic discourse dominant |

Figure 1: Emmerson's language grid (Emmerson 1986). The shaded area (Vickers 2006) represents the space occupied by auditory display.

## 4.1 Mimetic Discourse and Auditory Display

We need to recast our terms slightly to accommodate auditory display into Emmerson's language grid as auditory display is not (predominantly) a universe of musical practice. In Emmerson's scheme aural discourse refers to those pieces of music in which the composer focuses on the abstract discourse of interacting sounds and the patterns they form. Mimetic discourse is that in which, at its extreme, the composer is using recorded sounds to directly evoke images of the environment. Constrast this with aural discourse in which 'our perception remains relatively free of any directly evoked image' (Emmerson 1986: 19).

---

[8] In Emmerson's system an *abstract* syntax is one in which the musical ideas and sound materials being used are organized independently from each other, that is, where the structure of the piece has no intrinsic relationship to the timbres being used. An *abstracted* syntax is one in which the organization is abstracted directly from the sound-generating materials (Vickers 2006). For example, if the relationships between recorded environmental sounds were carried through into the organization of a *musique concrète* composition then we would say the music's syntax had been abstracted from the sonic materials. Thus, traditional music composition tends to use abstract syntax in which the notes on the score are related only to each other or to some external device (e.g., random numbers, the Fibonacci series, etc.).

However, in auditory display the use of sonic material is not a determinant of the type of discourse because sonification and audification practice employs abstract and concrete sounds alike. Regardless of the sonic material used, all auditory display is an exercise in signification and, hence, is designed to evoke concepts related to the material the sound represents. If we rename 'mimetic' to become 'representational' or 'indexical' we are closer to the true state of affairs as far as auditory display is concerned (see also Vickers & Hogg 2006).

Something (a gesture, an utterance, a sign, etc.) that is indexical points to (indicates) some other thing that is external (an entity, an idea, etc.).[9] In Peirce's semiotics an index is one of the three representational types to which a sign can belong, the other two being 'symbols' and 'icons'. In this scheme a symbolic sign is one that has a completely arbitrary relationship to the object it stands for: the association is conventional and must be learnt before it can be understood (e.g., traffic lights). An iconic sign is one that is perceived to resemble or imitate its referent object (e.g., a caricature). Finally, an indexical sign is directly connected to its object in a physical or causal sense. For example, smoke becomes an indexical sign of a fire (Chandler 2007). Therefore, auditory display assumes the semiotic meaning of indexicality, that is, a measure of the arbitrariness of a mapping. However, we must note that in auditory display, index, icon, and symbol are not necessarily discrete categories of sign but can be points along a continuum.

Thus, in the earcon-based approaches in which hierarchies of musical phrase structures are used to represent hierarchies of information (e.g., Brewster 1998) the association between musical motif and referent object is symbolic as it arbitrary and needs to be learnt. Auditory icons, as the name would suggest, are more iconic in their representation; take, for example, Gaver's auditory progress bar (Gaver 1986) in which the progress of a file being copied was mapped to the sound of a jug being filled with water. Thus, sonification tends to be symbolic or iconic with the indexical representations being found more in the domain of audification in which the sound heard is caused (more or less directly) by the data. Kramer set it out as a representational continuum from analgogic (indexical) to symbolic (Kramer 1994b). So, we say that indexicality in auditory display ranges from the purely indexical representations at one end to the symbolic representations at the other. Combining this with the mimesis-to-representation transformation discussed above we get Figure 2 which shows auditory display framed along the lines of Emmerson's language grid. We observe that auditory displays composed of mostly abstract syntax have lower indexicality than those using an abstracted syntax (e.g., Hayward's seismograms (Hayward 1994)).

---

[9] Indexicality is a concept from philosophy which is often used interchangeably with the linguistics term *deixis* and is also used in semiotic explanations of *sign*.

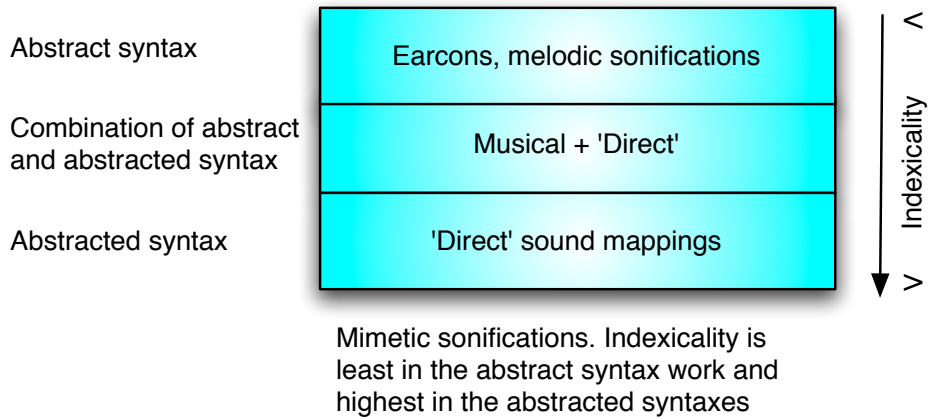

Figure 2: Indexicality in Auditory Display (adapted from Vickers 2006).

## 4.2 Shifting the Aesthetic Perspective

The Aesthetic Perspective Space, or APS, (Vickers & Hogg 2006) in Figure 3 offers a vehicle for talking about auditory display design from an aesthetic perspective. Auditory display sits alongside sonic art much like visualization can be discussed using the language of graphic design or even visual art. The APS focuses on organized sound as this offers the most mature set of aesthetics.

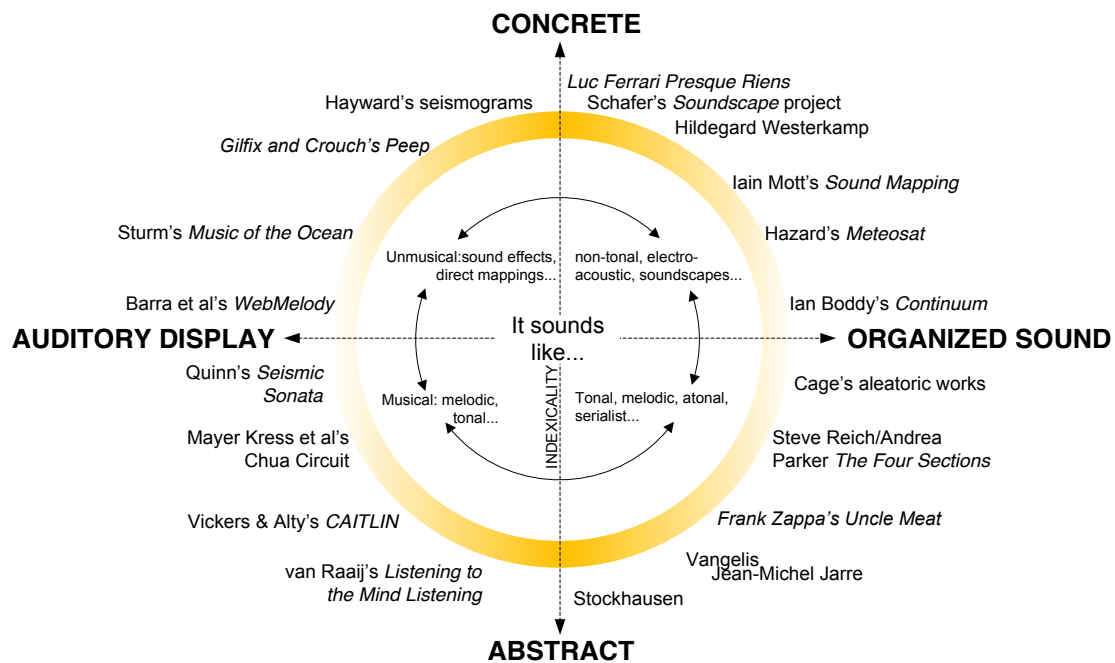

Figure 3: The Aesthetic Perspective Space (adapted from Vickers & Hogg 2006). Auditory display is ranged along the abstract/concrete continuum in the left half, whilst music and organized sound works are situated in the right half but along the same continuum.

The APS permits a kind of typecasting activity in which an auditory display is treated *as if it were* a work of organized sound. If it is viewed as organized sound then what kind is it? The APS is organized so that auditory displays lie in the left half and music on the right. Indexicality runs from north to south, being higher in the the northern half where concrete syntaxes predominate. Folding the APS along the indexicality axis enables this typecasting to take place. For example, if the display is based on tonal harmonies as in van Raaij's *Listening to the Mind Listening* (van Raaij 2004), then it would fold over onto the tonal music part of the APS.

AudioObject2.mp3 van Raaij's *Listening to the Mind Listening*
http://algoart.com/examples/VanRaaij.mp3

Likewise, auditory displays using more concrete sounds could be viewed as electroacoustic music. This is not to say that the sonifications are music in the sense that they have an artistic message at their core, but that they can be viewed as if they are music.[10] Then, once categorized, the prevailing aesthetic can be applied to the sonification design. The idea is that if, for example, a sonification is going to use tonal motifs for its communication, then if it obeys the aesthetic 'rules' (or semiotic code) for tonal music (whatever those rules are) then it will be more coherent and have more integrity than a sonification which has poor tonal musical design. Thus, the APS allows us to look at auditory display through the lens of organized sound and to explore the listening experience in those terms.

## 5 Ways of Listening

Hearing is a physical activity, a function of the human auditory system whereas listening is a mental or cognitive activity involving the mind. As Robert Jourdain put it, '*sound* (as opposed to *vibration*) is something that a mind *does*' (Jourdain 1997: xiv). Auditory display arises to solve problems in a variety of contexts (direct inspection, peripheral monitoring, etc.) and so the listening strategies adopted are not homogeneous. Schaeffer's *quatre écoutes* provide a good starting point for exploring the listening experiences encountered in auditory display. Table 1 sets out Schaeffer's *quatre écoutes*, his four modes, or ways, of listening (Schaeffer 1967).

|  | Abstract | Concrete |
|---|---|---|
| **Objective** | 4. *Comprendre* | 1. *Écouter* |
| **Subjective** | 3. *Entendre* | 2. *Ouïr* |

Table 1: Schaeffer's *quatre écoutes*, or four ways of listening.

Schaeffer divided listening between two sets of oppositions: abstract/concrete and objective/subjective which gives rise to the four modes shown in Table 1. The objective modes are concerned with the object of perception, that is the sound is deictic and the objects or events it points to are the focus of the listening experience. The two subjective modes are listener centred, the referent object is discarded, and the sound itself becomes the focus of the subject's attention (Kane 2007). The concrete modes are concerned with the sound without trying to extract any meaning from it (beyond its possible deictic function) whilst the abstract modes deal with situations in which we do abstract meaning from the sound. As we shall see in the discussion below, auditory display for the most part involves the listener in the objective modes.

---

[10] The primary intention of an auditory display is always to communicate, as unambiguously as possible, information about a data source. There may be a secondary artistic intention (see §5.4 for a discussion of such auditory displays) but this is subordinate to the primary goal of supporting specific meaning-making.

## 5.1 The Objective Modes: *Écouter* and *Comprendre*

*Écouter* is the act of listening to something through the agency of a sound, that is, we are attempting to identify its source. This is an example of Peirce's thirdness for the sound brings us into relationship with the thing that made or caused it and the sound is very much a sign in the semiotic sense. In auditory display terms we would be recognizing the mapping between data source and sound. For example, we might listen to an auditory display and remark 'that sound corresponds to the number of packets being received by our network'.

This listening mode is fundamental to auditory display and no auditory display can function without it. Even though our listening may fall into another mode (discussed below) we are always pulled back to this one at the point at which we try to decode the sound. In terms of auditory display it may be more proper to say no listening occurs without *Écouter*. It might be possible to let go of *écouter* in musical listening (see §5.3 below) but it is certainly not possible when listening to (or *reading*) an auditory display.

*Comprendre* is that mode in which having identified the referent object of the acoustic sign through *écouter*, we attribute meaning to (or infer meaning from) the sign. We have gone beyond *écouter*'s identification of source and instead are asking what the sound means. For example, having established that a sound corresponds to the number of packets being received on our network, *comprendre* lets us valorize the signal and perhaps remark that 'the number of packets on the network is high', or make an extended inference such as 'the packet count suggests we might be under a denial-of-service attack'.

## 5.2 The Subjective Modes: *Ouïr* and *Entendre*

If *écouter* and *comprendre* deal with referent objects, the subjective modes treat the sound as a subject in its own right. *Ouïr* concerns passive perception, it is a hearing in which we neither listen nor seek to understand and for that reason this mode is rarely encountered in auditory display. The closest we might get is in live monitoring situations in which the auditory display becomes part of the background, a kind of mapped Muzak if you will, which exists just below the attentional surface. However, we would switch out of this mode as soon as the display changes in a perceptually salient way causing us to notice a significant change in the state of the display.

*Entendre*, the selective hearing mode, lets us attend to aspects of the sound that attract our attention. Because this is a subjective mode we are not listening to what the attributes of the sound mean in terms of any external object but to how those attributes affect the sound itself. For instance, we might be interested in the unfolding timbre of the sound for its own sake. *Entendre* really has no place in listening to auditory displays in terms of their intended purpose. We might say that *écouter* and *comprendre* are functional auditory display listening modes, *ouïr* is pseudo-functional, and *entendre* is non-functional.

## 5.3 Auditory Display and the Acousmatic Reductions

One of Schaeffer's goals was to achieve a reduced listening (*écoute réduite*) in which the sound-in-itself becomes the self-referential object with which we are drawn into relation. His first reduction was to 'bracket out the spatio-temporal causes' (Kane 2007: 17) leaving acousmatic sound in which the causes behind it are rendered invisible. It is in this first acousmatic reduction that the objective listening modes above sit. The second reduction removes reference to anything beyond the sound giving us the two subjective listening modes which move us closer to Schaeffer's reduced listening experience in which any signification is lost.

Acousmatic listening occurs whenever the visible cause of the sound is removed, such as happens in radio broadcasts. On the face of it, auditory display is a clear example of acousmatic production: data or events are converted to sound, the user listens to the sounds and makes inferences about the state or values of the system or data. What could be simpler than that? However, scratch the surface and auditory display reveals itself to be a diverse collection of practices. Sometimes it is acousmatic, sometimes it is not. For example, audification is almost entirely acousmatic, parameter-mapping sonification is often acousmatic but doesn't have to be, and model-based sonification *can* be acousmatic but usually isn't.

Audification is acousmatic because it typically involves the playback of large series of data compressed into a short audio clip (e.g., several months of seismic data played back over tens of seconds, or a minute); there is no real interaction, save for possibly pausing, rewinding, and replaying the audification. Parameter-mapping sonification (PMS) is the set of techniques in which different data items, or system properties, are mapped to specific audio parameters. For example, in a stock market scenario technology stocks could be mapped to one timbre, bank stocks to another timbre, and so on. The spot price of each stock could be mapped to pitch and the trading volume might control the Q point of a band-pass filter that sweeps the sound. PMS is often non-interactive in that the system generates the audio and the user consumes it. The user might make adjustments to the system in response, but the paradigm is often one of information display (such as a chart projection in visualization). Of course, interactive parameter-mapped sonification systems do exist. For example, consider Vogt et al. (Vogt, Pirrò, Kobenz, Höldrich, & Eckel 2010) who used sonification in a system for assisting with motor control in a physiotherapy context. In model-based sonification (MBS) the data set controls a parameterized sound model, such as Hermann and Ritter's resonator (Hermann & Ritter 2004). The user interacts with the model and gains knowledge about the data from the sounds that come out. This is a highly interactive process and it is clearly not acousmatic for the sound we hear is a direct response to the stimuli we provide to the system.[11]

---

[11] For more on audification, PMS, and MBS see Dombois and Eckel (Dombois & Eckel 2011), Grond and Berger (Grond & Berger 2011) and Hermann (Hermann 2011) respectively.

**Acousmatic and Direct Listening**

So, it is clear that Schaeffer's objective modes apply to auditory display, but it is also clear that not all auditory displays involve acousmatic listening and even those that do support it do not necessarily do so all the time. If, for example, we were to build a sonified spreadsheet program in which the values of the cells and information about inter-cell relationships are mapped to an auditory display, then every time we edit the spreadsheet the sound would change. The cause of the sound is visible to us — we change a value in a cell and the auditory output changes accordingly.[12] But what if we altered *this* cell here and the sound of *that* cell over there changed because *this* cell was, unbeknownst to us, implicated in a function determining the value of the other cell? Is the sound now acousmatic even though we are the unwitting cause of the changed sound? Or what if we are away from our desk and somebody else edits the spreadsheet whilst our back is turned? It would seem that sometimes auditory displays are acousmatic but other times they are not. Let us consider some other examples. In a programming debugging scenario, a parameter-mapped sonification might be generated that tracks the run-time behaviour of a computer program. In this situation the sonification is acousmatic for the program is running inside the machine and we certainly cannot see its execution. But now imagine a sonically-enhanced program editor that lets us highlight a block of code and listen to its sonification; this would seem not to be acousmatic as the cause of the sound is present and available to us (albeit indirectly through the agency of the visual display).

Schaeffer distinguished between acousmatic (hidden-cause) listening and direct listening 'which is the "natural" situation where sound sources are present and visible' (Chion 1983: 11). Because auditory display operates in both the acousmatic and direct realms Schaeffer's objective listening modes are insufficient to capture this state of affairs. To our rescue come Chion's causal and semantic listening modes (Chion 1994) which sit astride the acousmatic/direct divide. Casual listening is very much like *écouter* except that it is not restricted to the acousmatic reduction: when the cause is invisible causal listening directs us to its source and when the cause is visible the 'sound can provide supplementary information about it' (Chion 1994: 25). Semantic listening has its counterpart in *comprendre* but also admits both acousmatic and direct presentation.

Gaver (Gaver 1989) offers us two listening modes specifically related to auditory display: everyday and musical listening. Gaver defines musical listening as the 'experience of hearing sounds *per se*' whilst 'hearing attributes of sound-producing events is … *everyday* listening'. Gaver continues:

> If the dimensions of musical listening correspond to fundamental physical attributes of the sound, the attributes of everyday listening correspond to the attributes of the source.

---

[12] There is an interesting question that the word 'visible' raises, for to a blind user the cause of the sound is not visible, but it is certainly immanent and directly perceptible.

Thus, we see that Gaver's everyday listening subsumes Schaeffer's objective modes and Chion's causal and semantic listening, whilst his musical listening encapsulates Schaeffer's subjective modes. Table 2 shows the original Schaefferian table augmented with Chion's and Gaver's modes.

|  | Abstract | Concrete |  |
| --- | --- | --- | --- |
| **Direct** | 8. ? | 5. ? | } **Objective** |
| **Acousmatic** | 4. *Comprendre* | 1. *Écouter* |  |
|  | 3. *Entendre* | 2. *Ouïr* | } **Subjective** |

$$\text{Musical} = \{2, 3\}, \text{Everyday} = \{1, 4, 5, 8\}$$
$$\text{Causal} = \{1, 5\}, \text{Semantic} = \{4, 8\}$$
$$\text{Direct} = \{5, 8\}, \text{Acousmatic} = \{1, 2, 3, 4\}$$
$$\text{Objective} = \{1, 4, 5, 8\}, \text{Subjective} = \{2, 3\}$$

Table 2: Modes of listening: Schaeffer, Chion, and Gaver. The direct listening space opens up the possibility of further listening modes mirroring those in the acousmatic space. This table shows the positions of the new modes 5 and 8.

We see from Table 2 that Schaeffer's *quatre écoutes* are atomic whilst those of Chion and Gaver are groups. We also note that that there is no explicit name given to the two modes (8 and 5) implied by the existence of direct listening. Let us call these modes *DAO* (direct abstract objective listening) and *DCO* (direct concrete objective listening) respectively. We also note that two more modes concerned with direct subjective listening are implied by the categorization of Table 2: direct abstract subjective (*DAS*) and direct concrete subjective (*DCS*). All eight such *écoutes* are shown in the lattice diagram of Figure 4. We can see that Gaver's everyday listening is synonymous with objective listening and musical listening with subjective listening. Chion's causal listening is the concrete-objective subset of modes and semantic listening is the abstract-objective subset.

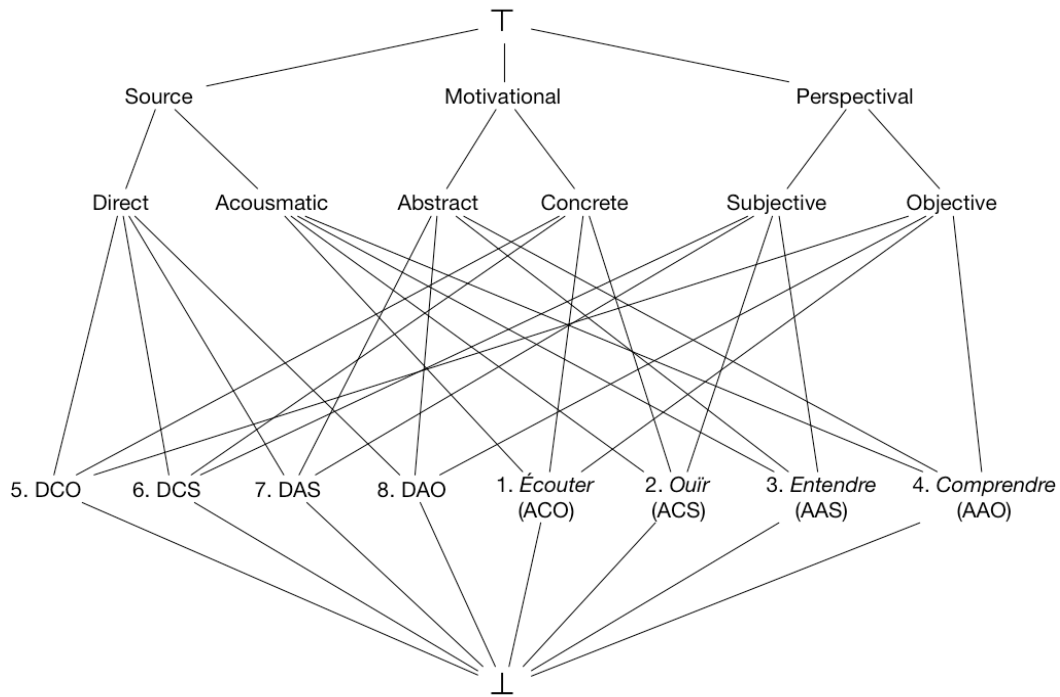

Figure 4: A lattice of listening modes. The modes are derived as combinations of the three pairs of oppositions that subdivide the universal type ⊤: Direct or Acousmatic; Abstract or Concrete; Subjective or Objective. At the bottom is the absurd type, ⊥, which is true of nothing (it is the the subtype of all the types in the hierarchy). In ontology every type hierarchy has a universal type and an absurd type and so they are both shown here for the sake of completeness.

## 5.4 The Emergent Modes

From Figure 4 we see there are four emergent atomic modes that arise from adding the direct/acousmatic opposition to the *quatre écoutes*. These eight modes are shown in tabular form in Table 3. The four emergent modes are the direct listening complements of Schaeffer's *quatre écoutes*. Where do they fit in the auditory display listening experience?

|  | Abstract | Concrete |  |
|---|---|---|---|
| Direct | 7. DAS | 6. DCS | Subjective |
|  | 8. DAO | 5. DCO | Objective |
| Acousmatic | 4. *Comprendre* | 1. *Écouter* | Objective |
|  | 3. *Entendre* | 2. *Ouïr* | Subjective |

Table 3: Eight atomic modes of listening.

DCO (the complement of *écouter* or ACO) as we have already seen is covered by Chion's causal listening and Gaver's everyday listening. In direct concrete objective listening we are treating the sound as an indexical (deictic) sign of a visible or present cause. This would occur when interacting with a model-based sonification or when a parameter-mapped display permits interaction.

DCS is like its complement *ouïr* (ACS) except that the cause is visible. It is difficult to imagine this as a functional listening mode for an auditory display for it is a passive hearing mode. What it might suggest is an unsuccessful interactive auditory display, perhaps one that has audio-visual output but where the auditory aspect is being ignored.

DAS (complement of *entendre*/AAS) is the fourth subjective mode that has little place in auditory display. Unlike DCS we are interested in the sound being made but for its own sake. This mode might apply when engaging with an interactive sonic art installation: the user makes some input, or gesture, to which the system responds and it is the sound that emanates which is of interest. The interaction serves only to modify the sound. This is, perhaps, a mode experienced by the player of a musical instrument (except that a musician might be expected to move into a more objective mode becoming interested in the feelings, moods, ideas, etc. that the music evokes or suggests).

DAO is the most common of the direct listening modes for auditory display. It is like its complement *comprendre* (AAO) but the cause of the sound is visible or present. This is the mode that would apply when using an interactive auditory display to explore a data set such as would be found in model-based sonification systems.

So, while some auditory displays support acousmatic listening and some do not we can go so far as to state that Schaeffer's reduced listening cannot apply to an auditory display as long as one is listening to it *as a display*. Sonification, as the spelling checker in my computer's word processor keeps insisting, is always an activity of signification so there is no such concept as sound-in-itself in auditory display for the meaning of the sound must always be outside itself. All sounds in an auditory display are purposeful and point to something important; the causal relationship is central, the mapping is key.

Reduced listening is possible only if the auditory *display* aspect is desired to be ignored or bracketed. This is of course feasible just as one can enjoy the experience of Beethoven's *Pastoral* symphony without having read the programme notes. Indeed, some works of auditory display were deliberately designed with this duality in mind. For example, take Sturm's *Music From the Ocean* (Sturm 2002), a CD of sonifications of data gathered from a set of Pacific Ocean buoys.

AudioObject3.mp3: Sturm's Pacific Ocean sonification
http://www.mat.ucsb.edu/~b.sturm/music/PacificPulseStereo.mp3

Sturm deliberately composed the sonifcations to function both as a useful information display and as music to be enjoyed in its own right. The sleeve notes contain detailed descriptions of the data-to-sound mappings, yet one can choose to completely ignore these *programme notes* and just enjoy the sonic experience. So, whilst these compositions of ocean-buoy data were expressly designed to allow the state of the Pacific ocean to be inferred through listening, they also serve as ambient compositions that can be listened to in their own right as pieces of music. Sturm's example serves to highlight that auditory displays that are designed with a more elaborate, even musical, aesthetic in mind can work on two levels: as communicators of information (in the scientific or information visualization sense) and as sonic art. In both cases the sound leads to sensuous perception: as an auditory display in terms of information inference through perception by the senses, as sound art in terms of connection with the sublime. However, whereas art aims to connect us with the sublime an auditory display is, in Manovich's language, 'anti-sublime' (Manovich 2002) because its goal is not to achieve outstanding artistic expression; for auditory display designers good aesthetic practice steers us towards achieving ease of use (Barrass & Vickers 2011).

Sturm is not alone in his sonifications-as-music. For example, see *inter alia*, Quinn's *The Seismic Sonata* (Quinn 2000),

AudioObject4.mp3: The Seismic Sonata
http://www.drsrl.com/tour/music/Track21.mp3

Polli, Van Knowe, and Vara.'s performance installation *Atmospheric/Weatherworks* (Polli, Van Knowe, & Vara 2002),

AudioObject5: Atmospherics/Weatherworks
http://www.andreapolli.com/studio/atmospherics/

all ten pieces presented at the ICAD 2004 *Listening to the Mind Listening* concert (Barrass & Vickers 2004),[13] or the ICAD 2006 *Global Music — The World by Ear* concert (de Campo 2006).[14] The key feature of these examples is that they are legitimate sonifications in addition to their status as sonic art. That is, once the sonification rules have been learnt and understood one is able to listen to the sonifications and gain measurable insight into the data they represent.[15] Auditory display, then, embraces the acousmatic but without the second reduction, it offers a form of non-speech acousmatic diegesis.

## 5.5 A New Listening Experience?

In his review of sonic art and post Schaefferian music Kim-Cohen (Kim-Cohen 2009) challenges the very concept of reduced listening and urges us to leave Schaeffer's sound-in-itself behind and embrace, instead a 'non-cochlear' sonic art. According to Kim-Cohen, a non-cochlear sonic art 'seeks to replace the solidity of the *objet sonore*, of sound-in-itself, with the discursiveness of a conceptual sonic practice' (p. 218). But, while we can view auditory display as organized sound it can never be a non-cochlear art form in its role as information display. For Kim-Cohen a non-cochlear sonic art 'does not accept the resolution of sound-in-itself — not because it seeks another kind of resolution, but because it denies the possibility of resolution, ipso facto' (p. 260). This is in direct opposition to auditory display where the express intention is to achieve resolution, for the listener to arrive at some verifiable knowledge of the referent object that the auditory display portrays. As Rosenblum so succinctly put it, 'we hear events, not sounds' (Rosenblum 2004: 220).

---

[13] http://www.icad.org/websiteV2.0/Conferences/ICAD2004/concert.htm
[14] http://www.dcs.qmul.ac.uk/research/imc/icad2006/concert.php
[15] Contrast this with data-driven art in which the goal is always artistic, the data serving only as a catalyst for the art.

## 5.6 Hierarchical Listening

In their exploration of sound design for user interfaces Tuuri, Mustonen, and Pirhonen (Tuuri, Mustonen, & Pirhonen 2007) took Huron's (Huron 2002) theory of auditory evoked emotion and recast it to identify eight different levels of listening for auditory display, namely reflexive, connotative, causal, empathetic, functional, semantic, critical, and reduced. Reflexive listening is a pre-conscious automatic reaction to sound (the sort of response an alarm might generate). Connotative listening is also pre-conscious but involves more signification than reflexive listening and accesses associations between the sound and the physical properties or cultural associations of the object or event that caused it. For example, a connotative response would be knowing that a vehicle has a large engine just from the sound it makes. The causal mode is more denotative and makes specific identification of the sound's cause (such as recognizing that a man just spoke or a telephone just rang) (Mustonen 2008). Empathetic listening focuses on perceiving the mental state of the source (where applicable). For instance, a slammed door might be interpreted as denoting anger in the person who slammed it. Functional listening focuses on the purpose of the sound in context. For example, in a sonification system one would recognize that a certain sound is confirming some user action (e.g., when a web browser makes a sound when a hyperlink is clicked). Semantic listening has broadly the same meaning as Chion's mode of the same name. Critical listening is concerned with reflective judgements, making decisions about the appropriateness of a sound in its context. Reduced listening is the same as Schaeffer's with the focus being on sound-in-itself, devoid of any signification.

Tuuri et al. (Tuuri et al. 2007) arranged these modes in a hierarchy of pre-conscious modes (reflexive and connotative), source-oriented modes (causal and empathetic), context-oriented modes (functional, semantic, and critical) and the single quality-oriented mode (reduced). When used to view auditory display they identify that earcons (a symbolic technique) require the use of two modes: the semantic in order to understand the coding scheme used in the earcon design and the reduced in order to hear the earcon as 'musical parameters' (p. 17). Auditory icons, on the other hand (an iconic representational device) use causal listening (to recognize the sound's source), connotative listening (to recognize the physical properties of the sound), and functional listening. However, it is not clear why they do not think causal listening also plays a part in earcon recognition as one needs to recognize the data object signified by the earcon in order to fully comprehend the signal. This hierarchy is also clearly not orthogonal to the eight discrete listening modes discussed above. Like Chion's and Gaver's modes, Tuuri et al's classification also sits astride the four direct and four acousmatic modes in Table 3. A helpful contribution is the perspective these modes offer, especially that of a functional mode of listening in which we explicitly attend to the role of a sound in an auditory display. This functional consideration alone serves to show how auditory display is fundamentally a utilitarian practice. Whilst Schaeffer's scheme does not capture this aspect well, we must not forget that functionalism is not the sole preserve of auditory display. Composers have long known the functional value of the *leitmotif* for signalling events and guiding the listener and this language has found its way into auditory display design. For example, the musical devices attached to programming language objects were called *leitmotifs* in my own program sonification system (see Vickers 1999).

## 6 Conclusions

While much has been written about listening in the domain of organized sound, listening has attracted relatively little interest in the field of auditory display which relies entirely upon the user possessing quite sophisticated listening skills. Literature dealing with perception, physiology, and psychoacoustics abounds, but these studies tend to focus on the human hearing response. We know quite well, for example, what the limits of human audition are, how perception of intensity and frequency are co-implicated, how temporal masking works, how auditory illusions can be generated, how audio streams at the perceptual level, and so forth. Listening, as we know, is quite different from hearing and it is evident that the user of an auditory display engages in several different types of listening. We saw above how auditory display can be viewed through the lens of organized sound and the listening experience discussed in organized sound's language. However, it is also clear that just like typecasting data types in a computer programming language results in some incompatibilities or loss of power, the match between organized sound's frames of reference and auditory display is not perfect. The ways of talking about listening to music assume certain standpoints regarding the source material that do not necessarily map well to auditory display. For example, when we talk of intention and intended meaning in a musical composition we do so knowing that much of the dialogue takes place between the music and the listener without reference to the composer and that whatever meaning the listener constructs is, to some extent, valid, even if the composer's express intentions are never realized. This is in complete contrast to auditory display which, like its visualization sibling, stands or falls on its ability to communicate specific information as unambiguously as possible. The intended meaning in an auditory display is paramount and trumps all aesthetic sensibilities. And yet it has also been stated that successful auditory displays need aesthetic integrity (Barrass & Vickers 2011).

The analysis of auditory display in terms of Schaeffer's *quatre écoutes* has, at the very least, shown areas in which auditory display and organized sound are similar and where they differ. We see that auditory display requires a focus on objective listening modes, but that it also operates in both the acousmatic and direct spaces. The discussion also identified the existence of four discrete direct listening modes that mirror Schaeffer's four acousmatic modes. Regardless of any strides made in discussing the subjective listening experience in terms of, say, spectromorphologies, the work of auditory display will not be advanced unless it can be shown that such a focus on the *objet sonore* can also be leveraged in the very pragmatic sound design that auditory display requires.

I have tried above to cast a light on the auditory display listening experience using the language of organized sound. In fact, to say 'the' language of organized sound is to miss the point somewhat for I am aware that I have framed the discussion only in terms of Emmerson and Schaeffer (and by extension, Chion) and that other tools are available. The choice of language is important though for any language brings with it not only its syntax but also its paradigm and it is at this level that we get connotative differences. Take sonification. The technologist would talk of this in terms of representations, sound generators, loud speaker arrays, and so forth; the metaphor theorist would explain it in the language of source domains and target domains, using metaphors to encapsulate the meaning to be transferred; a computer scientist using the language of conceptual mathematics might describe the process as a creation of maps between the data object and the representational object, the inverses and compositions of the maps, and the algebra that can be used to reason about them; the sonification designer (if such an identifiable profession yet exists) might explain it in terms of the linkages between data dimensions and auditory or sonic-model parameters; the musician or even the sonic artist might explain it in yet quite different terms. What these different views describe, however, is the process of making perceptible to the senses that which inherently cannot be tangibly prehended.

The language they use determines how they reason about the process, but they are all talking about the same process, that of transforming data into sound, of making perceptible abstractions held in silicon. The above analysis has shown that using the language of organized sound has taken us some way to being able to describe more fully the various listening experiences that are involved in auditory display. Tuuri et al. showed how the language of auditory emotion could be used to discuss the auditory display listening experience. Again, they (like other auditory display researchers that have gone before) have borrowed the language of Schaeffer and Chion and have moulded it to fit this thing that looks and feels a lot like organized sound but which is ontologically different. Even Gaver's musical and everyday listening can be mapped onto Schaeffer and Chion, although his categories do push the borders around a bit, the better to fit the lie of the land.

Sometimes the scientist and the artist do the same things but do not recognize it. For example, it was commonplace in the 1960s and '70s for computer programmers to tune AM radios to pick up the interference from their mainframe computers because they found they could use these signals to diagnose faults in their programs as they ran. Then in 2003 Christian Kubisch initiated her *Electrical Walks* series in which participants would wander through cities equipped with headphones that rendered audible the electrical noise that was all around.[16] One was able to listen to the hidden activity of street lights, cash machines, transportation networks, and so on. This is, of course, what an auditory display researcher would recognize as audification, but the goal here was aesthetic. Coming from an artistic standpoint Kim-Cohen criticized the practice saying that 'to "read" the work as if it is conveying a message — as if it is the product of a legible intention — seems forced' (Kim-Cohen 2009: 111). Yet this is precisely what those programmers fifty years previously were doing, they were reading a signal that was never intended to be heard as a message and it conveyed real meaning that allowed real work to be done.

---

[16] See http://www.christinakubisch.de/english/install_induktion.htm.

What is needed is a language, a paradigm, the syntagma of which will enable us to capture the richness of auditory display in all its diversity and reason about the listening experience in clear terms from which will come greater understanding of the field. From understanding will come the knowledge of how to build better auditory displays. Psychology and perceptual science have such languages for exploring hearing psychophysical response. From this experiments can be designed which allow for hypotheses to be tested. Landy called for 'more methodologically strong evidence-based research' for underpinning sound-based music studies (Landy 2007: 223), and the same is true for the auditory display listening experience.